\documentclass[12pt]{article}
\usepackage[margin=1in]{geometry}
\usepackage{pdflscape} 

\usepackage{xr-hyper}

 \usepackage[hidelinks]{hyperref}

\usepackage{graphicx,verbatim,array,multicol, subfigure, color, lscape, mathrsfs,natbib, hyperref,enumitem}
\usepackage{psfrag, amsmath, amsfonts, epsfig, fancybox, setspace,soul,amsthm,amssymb}
\usepackage{bm} 

\def\supg{^{(g)}}

\def\cov{\mbox{cov}}

\def\supone{^{(1)}}
\def\supzero{^{(0)}}

\definecolor{green}{RGB}{000,150,100}
\definecolor{purple}{RGB}{150,000,180}

\newcommand{\jbwhat}[1]{\widehat{#1}}

\def\wDelt{\widehat{\Delta}}

\def\To{\rightarrow}
\newcommand{\var}{\mbox{Var}}
\newcommand{\wtilde}[1]{\widetilde{#1}}
\newcommand{\qmq}[1]{\quad\mbox{#1}\quad}

\begin{document}
\thispagestyle{empty}
\begin{center}
\vspace*{25mm}
\LARGE Group Sequential Testing of a Treatment Effect Using a Surrogate Marker\\
\vspace*{15mm}
\large Layla Parast and Jay Bartroff \\
\vspace*{5mm}
\large Department of Statistics and Data Sciences\\
The University of Texas at Austin \\
2317 Speedway STOP D9800, Austin, TX, 78712\\
\end{center}

\clearpage
\begin{abstract}
The identification of surrogate markers is motivated by their potential to  make decisions sooner about a treatment effect. However, few methods have been developed to actually use a surrogate marker to test for a treatment effect in a future study. Most existing methods consider combining surrogate marker and primary outcome information to test for a treatment effect, rely on fully parametric methods where strict parametric assumptions are made about the relationship between the surrogate and the outcome, and/or assume the surrogate marker is measured at only a single time point. Recent work has proposed a nonparametric test for a treatment effect using only surrogate marker information measured at a single time point by borrowing information learned from a prior study where both the surrogate and primary outcome were measured. In this paper, we utilize this nonparametric test and propose group sequential procedures that allow for early stopping of treatment effect testing in a setting where the surrogate marker is measured repeatedly over time. We derive the properties of the correlated surrogate-based nonparametric test statistics at multiple time points and compute stopping boundaries that allow for early stopping for a significant treatment effect, or for futility. We examine the performance of our testing procedure using a simulation study and illustrate the method using data from two distinct AIDS clinical trials.

\thispagestyle{empty}
\end{abstract}

\clearpage
\setcounter{page}{1}
\section{Introduction}

\begin{figure}[htbp]
\begin{center}
\includegraphics[scale=0.6]{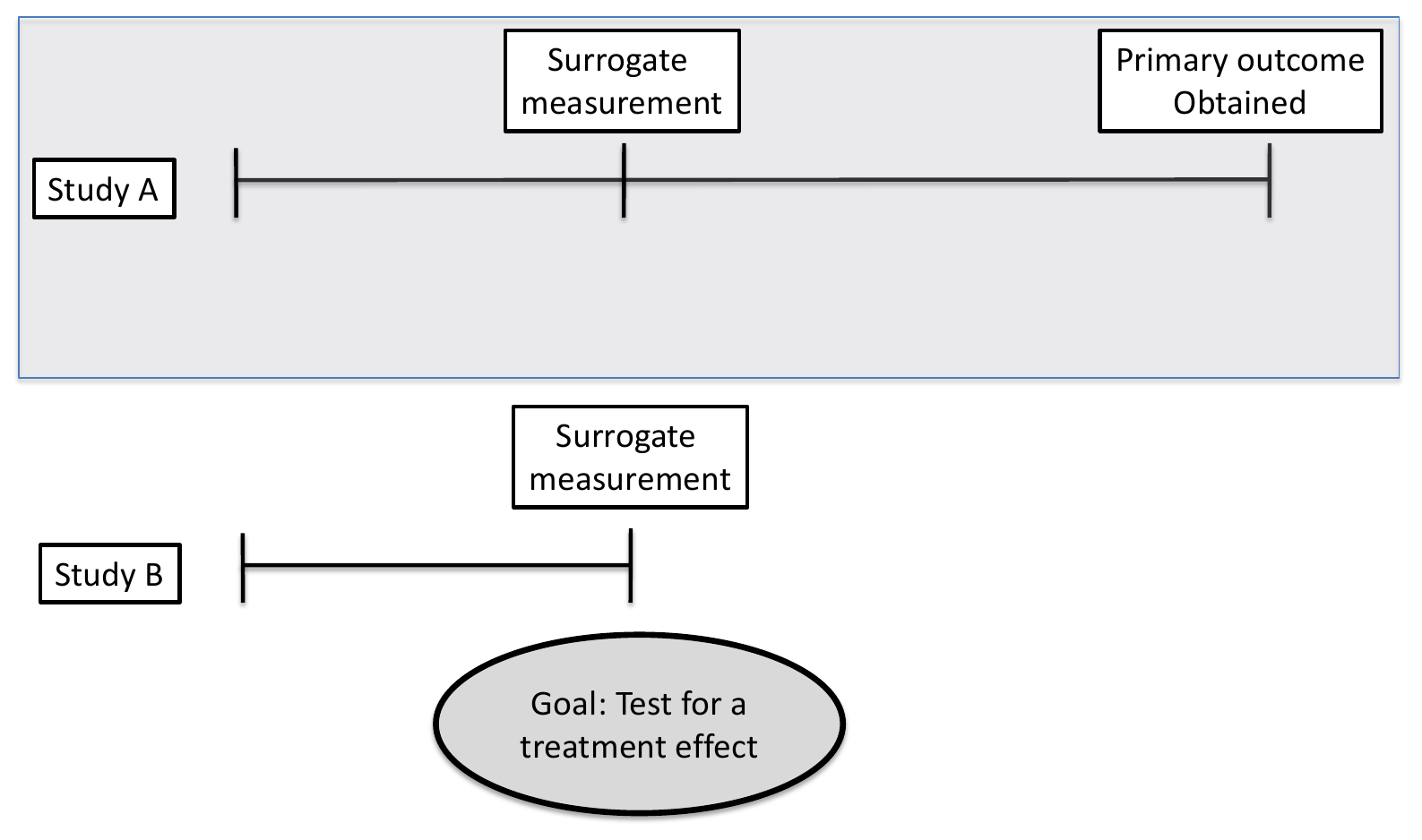}
\end{center}
\caption{Study A and B: Evaluating a Surrogate in Study A and Using the Surrogate in Study B to Test for a Treatment Effect}\label{pics}
\end{figure}

The ultimate promise of surrogate markers is that if they can be identified, then they can be used to make decisions about a treatment sooner. A surrogate marker is a measurement that replaces a primary outcome and is expected to predict the effect of a treatment \citep{temple1999surrogate,FDA_accelerated}. In studies where the primary outcome measurement necessitates long follow-up or is invasive or expensive, surrogate markers may lead to more timely decisions \citep{machado1990use,FDA_accelerated2}.  Rigorous statistical methods have been developed to identify valid surrogate markers \citep{elliott2023surrogate}. However, far fewer methods have been developed to actually use a surrogate marker to test for a treatment effect in a future study, which as mentioned above, is the ultimate goal. This is a particularly difficult problem to consider when the surrogate marker is not perfect e.g., it may not capture the entire treatment effect on the primary outcome. This concept is shown in Figure \ref{pics} such that there is a prior study, Study A, where the validity of the earlier (or less expensive) surrogate marker replacing the primary (longer term or more expensive) outcome has been examined. Interest lies in conducting a future study, Study B, and using only the surrogate marker information to test for a treatment effect.

Certainly, there is extensive previous work on utilizing a surrogate marker {\em in combination} with the primary outcome to test for a treatment effect. For example, \citet{li2022using} propose an approach to use both surrogate marker and outcome information to adaptively calculate conditional power in a group sequential trial. When both the primary outcome and surrogate marker are time-to-event outcomes, \citet{cook1996incorporating} and \citet{lin1991nonparametric} propose a weighted global test statistic in a group sequential trial allowing for the evaluation of the utility of the potential surrogate endpoint. Under the assumption that the surrogate and outcome are bivariate normal, \citet{tang1989design} explicitly demonstrates the advantage of a group sequential test using both endpoints in terms of reduced needed sample size. Similarly within a bivariate normal framework, \citet{anderer2022adaptive} develop an approach to combine the outcome and surrogate in a Bayesian adaptive design context.

However, in many settings, the question is not how to combine the surrogate marker and the primary outcome. Instead, the aim is to understand how to test for a treatment effect with the surrogate marker measurements \textit{only}, and thus avoid measuring the primary outcome. Methods that do address this question tend to either rely on fully parametric methods where strict parametric assumptions are made about the relationship between the surrogate and the outcome or assume the surrogate marker is measured at only a single time point.  For example, \citet{price2018estimation} propose to test for a treatment effect using a defined optimal surrogate at a single time point  which aims to predict the primary outcome with estimation via the super-learner and targeted super-learner. \citet{quan2023utilization} and \citet{saint2019predictive} proposed methods that use prior information about the treatment effect on a surrogate to plan a future study under the assumption that the true effects on the surrogate and primary outcome are bivariate normal.

Recent work \citep{parast2019using,parast2022using} has proposed model-free procedures to test for a treatment effect using only surrogate marker information measured in the future study (Study B) by borrowing information learned from the prior study (Study A). \citet{parast2022using} propose a kernel-based test statistic that is calculated using surrogate marker measurements from Study B obtained at a single, earlier time point. However, in practice the surrogate marker is often measured repeatedly over time (e.g., every 6 months) during the course of the study and thus, there is significant interest in applying sequential and group sequential testing methodology \citep{Bartroff13,Jennison00} to the surrogate setting. In this paper, we build upon the kernel-based testing framework by proposing group sequential procedures that allow for early stopping to declare efficacy, and a version that also allows for the possibility of futility stopping as well, i.e., early stopping to declare failure to reject the null hypothesis of no effect. The use of the longitudinal surrogate in the test statistics prevents them from having the independent increment structure which simplifies the design of many group sequential procedures \citep[see][]{Kim20,Spiessens00}. Instead, we compute the correlation structure of the  surrogate-based nonparametric test statistics and estimate it using the Study~A data. The estimates of the correlation structure are then used to compute the group sequential procedures' stopping boundaries.  We examine the performance of our procedures using a simulation study and illustrate the method using data from two distinct AIDS clinical trials.

\vspace*{-5mm}
\section{Setting, Notation, and Existing Approach}
\subsection{Setting and Notation}
\begin{figure}[htbp]
\begin{center}
\includegraphics[scale=0.6]{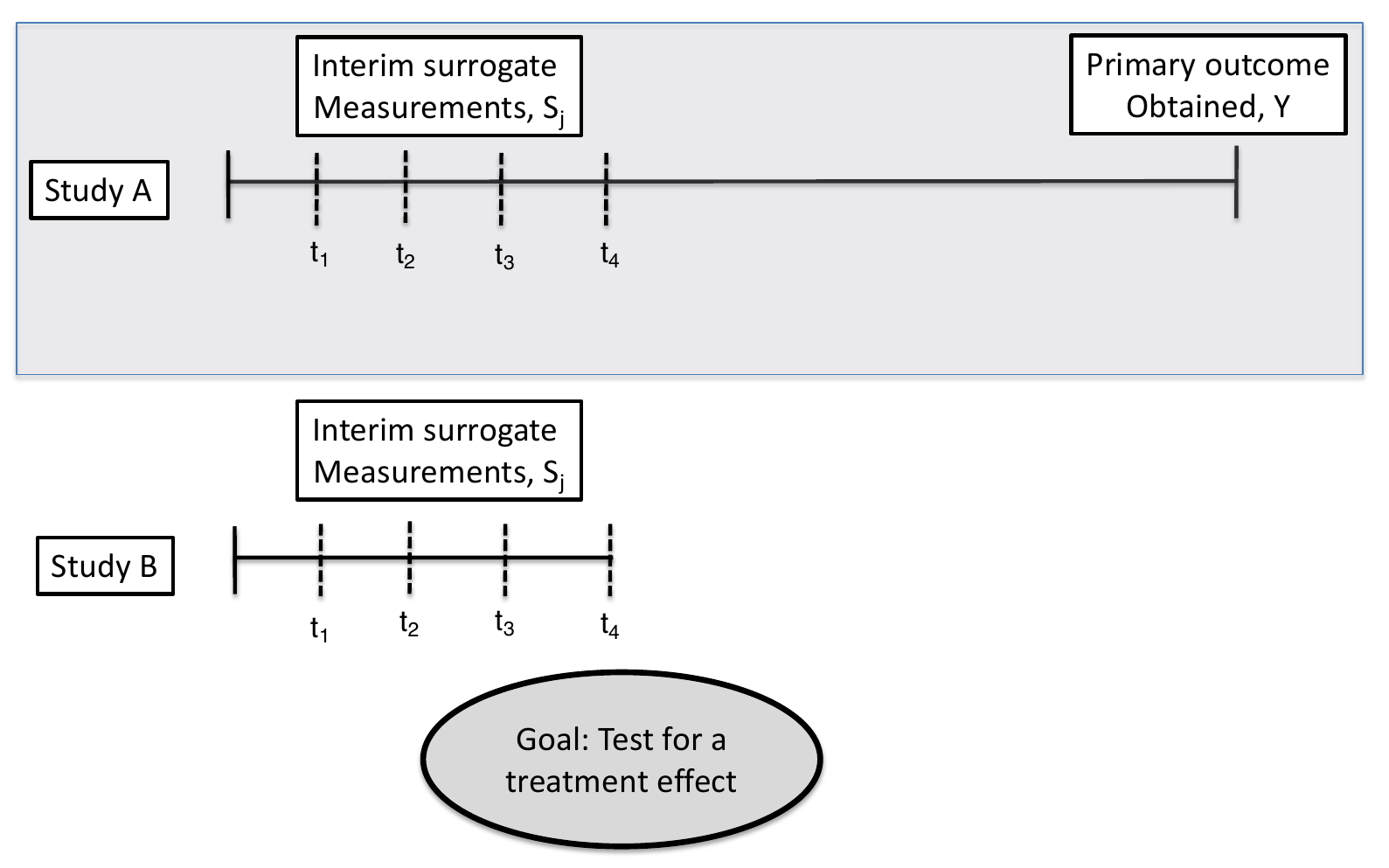}
\end{center}
\caption{Study A and B: Sequential Surrogate Setting}\label{pics_small2}
\end{figure}

In Figure \ref{pics_small2} we expand on Figure \ref{pics} to illustrate our setting of interest which has surrogate marker measurements over time. Let $Y$ denote the continuous primary outcome measured at study completion in Study A and let $S_j$ denote a continuous surrogate marker which is measured at multiple time points, $t_j, j=1,...,J$, during the study. Without loss of generality, we assume $Y \geq 0$. Let $G$ denote the treatment indicator where treatment is randomized and $G \in \{0,1\}$ (i.e., treatment vs. control). Our aim is to use information learned in Study A about the relationship between the primary outcome, the surrogate marker measurements, and the treatment to test for a treatment effect at an earlier time point in Study B, such that the duration of follow-up needed for Study B can be shortened and the primary outcome does not have to be measured in Study B. We use a subscript $L$ to explicitly denote the study and use potential outcomes notation where each person in Study L has a potential $\{Y_L \supone, Y_L \supzero, S_{jL} \supone, S_{jL} \supzero\}$ where $Y_L \supone$ is the outcome under treatment, $Y_L \supzero$ is the outcome under control, $S_{jL} \supone$ is the surrogate at time $t_j$ under treatment, and $S_{jL} \supone$ is the surrogate at time $t_j$ under control in Study $L$. For individual $i$ in treatment group $g$ in Study $L$, the \textit{observed} surrogate at time $t_j$ is $S_{Lgji}$. For individual $i$ in treatment group $g$ in Study $A$, the \textit{observed} outcome will be denoted as $Y_{Agi}$; the outcome in Study B is never measured. Let $n_{Lg}$ denote the number of individuals in treatment group $g$ in Study $L$, and $n_L = n_{L0} + n_{L1}$ where $n_{Lg}/n_L > 0$.  When feasible, for ease, we will drop the subscript $L$ from notation.  

Our primary goal is to test for a treatment effect on the primary outcome in Study B quantified as $ \Delta \equiv E(Y_B \supone - Y_B \supzero) = 0,$  \textit{without} measuring the primary outcome, $Y$, in Study B. In the following sections we describe an existing kernel-based test for a treatment effect in Study B using a surrogate measured at only a single time $t_j$ \citep{parast2022using} and possible naive approaches to sequential testing (Section \ref{single}), and then propose a novel approach to sequentially test for a treatment effect in Study B, when the surrogate is measured repeated over time (Section \ref{seqsection}). In Section \ref{assumptions} we detail our needed assumptions as well as methods to empirically examine these assumptions.

\subsection{Existing Approach\label{single}}
Suppose there is only a single time point~$t_j$, i.e., that the surrogate marker is measured at only one time point after randomization, but \textit{before} the end of the study. Here, we drop the subscript $j$ from $S$ as it is unnecessary, and denote the potential surrogate as $S_L \supg$ and the observed surrogate for person $i$ as $S_{Lgi}$, $L=A,B$ and $g=0,1$. The goal is to take advantage of information from Study A to test $H_0$ using the surrogate marker measurement only from Study B.  To achieve this goal we first consider using the testing procedure of \citet{parast2022using}, described below.  Note that $\Delta$ can be expressed as: 
\begin{eqnarray*}
\Delta &=& E(Y_B \supone) - E(Y_B \supzero) = \int \mu_{B1}(s)dF_{B1}(s) - \int \mu_{B0}(s)dF_{B0}(s)
\end{eqnarray*}
where $\mu_{Lg}(s) = E( Y_L \supg | S_{L} \supg=s)$ and $F_{Bg}(s)$ is the cumulative distribution function of $S_{B} \supg$. Of course, this expression involves $Y_B\supg$. \citet{parast2022using} suggest to instead focus on 
\begin{equation}
\Delta^* \equiv \int \mu_{A0}(s)dF_{B1}(s) - \int \mu_{A0}(s)dF_{B0}(s) \label{deltastar}
\end{equation}
and they refer to $\Delta^*$ as an \textit{earlier} treatment effect because it is defined before the end of Study B using only 1) the conditional mean, $\mu_{A0}(s)$, from Study A, and 2) the surrogate marker in Study B i.e., $Y_B\supg$ does not appear in $\Delta^*$. The motivation for the exact construction of $\Delta^*$ is as follows. The troublesome components of $\Delta$, in terms of trying to examine the treatment effect earlier, are $\mu_{B0}(s)$ and $\mu_{B1}(s)$ because these involve $Y_B\supg$, which is not measure until the end of Study B. Suppose we simply decided to replace these components with their parallel components from Study A, i.e., define $\Delta^* $ as $\int \mu_{A1}(s)dF_{B1}(s) - \int \mu_{A0}(s)dF_{B0}(s)$. It is true that this construction could still be referred to as an earlier treatment effect because it is defined without $Y_B\supg$. However, there are two problems with this construction. First, we would have to assume that $\mu_{B0}(s) = \mu_{A0}(s)$ and $\mu_{B1}(s) = \mu_{A1}(s)$; that is, that the conditional mean in each treatment group is transportable between the two studies. Second, it can be shown that unless $S$ is a perfect surrogate marker, it is possible that $\Delta^* > \Delta$ and in an extreme case where $S$ is a poor surrogate, $\Delta^*$ may erroneously indicate a treatment effect when in fact, $\Delta=0$. It is this second problem that is particularly worrisome. It is never the case in practice that we have a perfect surrogate, and the risk of erroneously concluding there is a treatment effect using the surrogate when there is not a treatment effect on the outcome is not a risk that we are willing to take. Therefore, we specifically want a  $\Delta^*$ that can give us some guarantee, for example, that $\Delta^*$ is a lower bound on $\Delta$. The definition in (\ref{deltastar}) does guarantee this property. In Web Appendix A, we show that under our assumptions (see Section \ref{assumptions}), $\Delta^* \leq \Delta$ and that when $\Delta =0 \Rightarrow \Delta^*=0$. In addition, we show that a valid test of the null hypothesis that $\Delta^*=0$ would also result in a valid test of the null hypothesis that $\Delta=0$ in the sense that the type 1 error rate $\leq \alpha$. Thus, to test the null hypothesis that $\Delta =0$ one could instead consider testing the null hypothesis that $\Delta^* =0$. Notably, $\Delta^*$ still requires a form of transportability between the two studies:  $\mu_{B0}(s) =\mu_{A0}(s)$ (see Section \ref{assumptions}), though this is less strong than requiring this equality in \textit{both} treatment groups.

In practice, $\mu_{A0}(s)$ is unknown and furthermore, Study A data are fixed (i.e., $n_A$ is fixed and does not $\rightarrow \infty$) in this testing framework. Therefore, this earlier treatment effect must be defined in a way that acknowledges and makes explicit this reliance on Study A, specifically:   
\begin{equation}\label{DE.1st.def}
    \Delta_E = \int \widehat{\mu}_{A0}(s)dF_{B1}(s) - \int \widehat{\mu}_{A0}(s)dF_{B0}(s).
\end{equation}
\noindent where $\widehat{\mu}_{A0}(s)$ is a consistent estimate of $\mu_{A0}(s)$,
\begin{equation}\label{mu.hat.FSS}
\widehat{\mu}_{A0}(s) = \frac{\sum_{i=1}^{n_{A0}} K_{h}(S_{A0i} - s)Y_{A0i}}{\sum_{i=1}^{n_{A0}} K_{h}(S_{A0i} - s)}
\end{equation}

\noindent and $\Delta_E$ is estimated by 
\begin{equation}\label{Del.hat.Etj}
    \widehat{\Delta}_E = n_{B1}^{-1} \sum_{i=1}^{n_{B1}} \widehat{\mu}_{A0}(S_{B1i}) -  n_{B0}^{-1} \sum_{i=1}^{n_{B0}} \widehat{\mu}_{A0}(S_{B0i}).
\end{equation} 
Here, $K(\cdot)$ is a smooth symmetric density function with finite support (e.g., standard normal density), $K_h(\cdot) = K(\cdot/h)/h$, and $h$ is a specified bandwidth which may be data dependent. Note that this estimate only uses $S$ data from Study B (no $Y$ data from Study B is used or assumed to be measured) and $S,Y$ data from Study A in group $Z=0$ only. To test the null hypothesis, $\Delta = 0$, \citet{parast2022using} suggest testing the null hypothesis: $\Delta_{E} = 0$, (based on the properties above regarding $\Delta^*$) using the Wald-type test statistic $W_{E} = \sqrt{n_B}\widehat{\Delta}_{E}/\widehat{\sigma}_{E}$.  The test rejects the null hypothesis when $|W_E| > \Phi^{-1}(1-\alpha/2)$. 
In Web Appendix B, we provide the form of $\widehat{\sigma}_{E}$ and discuss the asymptotic properties of $\widehat{\Delta}_{E}$.

To use this existing test in our setting in Figure \ref{pics_small2}, one could consider simply testing $\Delta_E$ using $W_E$ at each time $t_j$, but this would have a high Type 1 error. One could instead consider utilizing a Bonferroni correction where the null hypothesis at each time $t_j$ is rejected if the $W_E$ calculated at $t_j$ is such that $|W_E| > \Phi^{-1}(1-\alpha/2K)$ which we expect would be conservative. In the next section, we propose a compromise between these two approaches via group sequential procedures.

\section{Group sequential procedures} \label{seqsection}
Here, we propose a group sequential procedure that allows for early testing of a treatment effect, derive the properties of the correlated surrogate-based nonparametric test statistics at multiple time points, and develop stopping boundaries that allow for early stopping for a significant treatment effect or for futility.
We assume that in Study A we have measured the surrogate marker at time points~$t_j$, $j=1,\ldots, J$, where $J\ge 2$ is the maximum number of interim analyses, and the same time points will be utilized for measuring the surrogate marker in Study~B.  Throughout this section let $[J]:=\{1,2,\ldots,J\}$.

\vspace*{-5mm}

\subsection{Group sequential statistic}

We will utilize the estimators~\eqref{mu.hat.FSS}-\eqref{Del.hat.Etj} but repeatedly calculated with data from each time point~$t_j$. Thus, we first adjust the notation to reflect this.  For $j\in[J]$, recall that $S_{Lgij}$  is the Study~$L$ ($L=A,B$) surrogate for the $i$th patient in group~$g$ ($g=0,1$) measured at time~$t_j$ and $Y_{Agi}$ is the Study~$A$ outcome for the $i$th patient in group~$g$ ($g=0,1$); note that we never measure the outcome in Study B. For each $t_j$, for any $s$ within the support of $S_{A0ij}$, we define
\begin{equation}\label{mu.hat.j}
\widehat{\mu}_{A0j}(s) = \frac{\sum_{i=1}^{n_{A0}} K_{h}(S_{A0ij} - s)Y_{A0i}}{\sum_{i=1}^{n_{A0}} K_{h}(S_{A0ij} - s)}
\end{equation} which is the kernel estimator using Study~A data from time~$t_j$. In addition, let
\begin{equation}\label{DeltEJ}
\wDelt_{Ej} = \frac{1}{n_{B1}} \sum_{i=1}^{n_{B1}} \widehat{\mu}_{A0j}(S_{B1ij}) - \frac{1}{n_{B0}} \sum_{i=1}^{n_{B0}} \widehat{\mu}_{A0j}(S_{B0ij})
\end{equation} be the estimator of $\Delta_E$ at time~$t_j$. Let $\bm{\Delta}_E=(\Delta_{E1}, \Delta_{E2}, \ldots, \Delta_{EJ})^T$ where 
$$\Delta_{Ej} = \int \widehat{\mu}_{A0j}(s)dF_{B1j}(s) - \int \widehat{\mu}_{A0j}(s)dF_{B0j}(s),$$ 
$\mu_{Agj}(s) = E( Y_A \supg | S_{Aj} \supg)$, and $F_{Bgj}(s)$ is the cumulative distribution function of $S_{Bgj}$. To reflect the group sequential nature of this set up we replace the null hypothesis, $\Delta_E=0$ by
\begin{equation}\label{hoe_j}
H_E: \bm{\Delta}_{E} = \bm{0},
\end{equation}
the latter denoting the $J$-dimensional zero vector.  We test \eqref{hoe_j} via group sequential monitoring of the longitudinal process~$\jbwhat{\bm{\Delta}}_E=(\wDelt_{E1}, \wDelt_{E2}, \ldots, \wDelt_{EJ})^T$.

\subsection{Covariance and limiting distribution}
Conditional on $\{\widehat{\mu}_{A0j}(s)\}_{j\in[J]}$, it can be shown that as $n_{B0}, n_{B1}\To\infty$,  the suitably-normalized distribution of $(\jbwhat{\bm{\Delta}}_E-\bm{\Delta}_E)$ approaches a $J$-dimensional multivariate normal random vector with mean zero (see Web Appendix B), and we recall that $\bm{\Delta}_E$ is the zero vector under the null hypothesis~\eqref{hoe_j}.   For $j,j'\in [J]$, the covariance matrix of $\jbwhat{\bm{\Delta}}_E$ has elements 
\begin{multline}
\sigma_{jj'}:= \cov(\wDelt_{Ej},\wDelt_{Ej'}) =\cov\left(\frac{1}{n_{B1}} \sum_{i=1}^{n_{B1}} \widehat{\mu}_{A0j}(S_{B1ij}) - \frac{1}{n_{B0}} \sum_{i=1}^{n_{B0}} \widehat{\mu}_{A0j}(S_{B0ij}), \right.\\
\left.\frac{1}{n_{B1}} \sum_{i=1}^{n_{B1}} \widehat{\mu}_{A0j'}(S_{B1ij'}) - \frac{1}{n_{B0}} \sum_{i=1}^{n_{B0}} \widehat{\mu}_{A0j'}(S_{B0ij'})\right)\\
=\frac{1}{n_{B1}^2}\cov\left(\sum_{i=1}^{n_{B1}} \widehat{\mu}_{A0j}(S_{B1ij}), \sum_{i=1}^{n_{B1}} \widehat{\mu}_{A0j'}(S_{B1ij'})\right) + \frac{1}{n_{B0}^2}\cov\left(\sum_{i=1}^{n_{B0}} \widehat{\mu}_{A0j}(S_{B0ij}), \sum_{i=1}^{n_{B0}} \widehat{\mu}_{A0j'}(S_{B0ij'})\right)\\
=\frac{1}{n_{B1}^2}\sum_{i=1}^{n_{B1}}\cov\left( \widehat{\mu}_{A0j}(S_{B1ij}),  \widehat{\mu}_{A0j'}(S_{B1ij'})\right) + \frac{1}{n_{B0}^2} \sum_{i=1}^{n_{B0}} \cov\left(\widehat{\mu}_{A0j}(S_{B0ij}), \widehat{\mu}_{A0j'}(S_{B0ij'})\right)\\
=\frac{1}{n_{B1}}\cov\left( \widehat{\mu}_{A0j}(S_{B11j}),  \widehat{\mu}_{A0j'}(S_{B11j'})\right) + \frac{1}{n_{B0}} \cov\left(\widehat{\mu}_{A0j}(S_{B01j}), \widehat{\mu}_{A0j'}(S_{B01j'})\right).\label{cov.Dj.Dj'.stat}
\end{multline}
Importantly, \eqref{cov.Dj.Dj'.stat} explicitly relies on Study A being fixed, which is our setting; we discuss this more in Section \ref{discussion}. Taking $j=j'$ in \eqref{cov.Dj.Dj'.stat} 
gives the variance
\begin{equation}\label{var.Dh.j}
\sigma_{jj}=\var\left( \wDelt_{Ej}\right)=\frac{1}{n_{B1}}\var\left( \widehat{\mu}_{A0j}(S_{B11j})\right) + \frac{1}{n_{B0}} \var\left(\widehat{\mu}_{A0j}(S_{B01j})\right).
\end{equation}

Let $\bm{D}$ be the $J\times J$ diagonal matrix with entries $\sigma_{11}^{-1/2}, \sigma_{22}^{-1/2},\ldots, \sigma_{JJ}^{-1/2}$, and let $\bm{\Sigma}=\cov(\bm{D}\jbwhat{\bm{\Delta}}_E)$ which is the $J\times J$ covariance matrix with $(j,j')$ entry equal to $\sigma_{jj'}/\sqrt{\sigma_{jj}\sigma_{j'j'}}$ and, consequently, diagonal entries equal to $1$. The null case of the multivariate normal limit mentioned above is that, under \eqref{hoe_j},
\begin{equation}\label{D.hat.lim}
    \bm{\Sigma}^{-1/2}\bm{D}\jbwhat{\bm{\Delta}}_E\To N_J(\bm{0}, \bm{I})
\end{equation} as the $n_{Bg}\To\infty$, where $\bm{I}$ is the $J\times J$ identity matrix. Although $\jbwhat{\bm{\Delta}}_E$ has the multivariate normal limit~\eqref{D.hat.lim}, it does \emph{not} have the independent increment structure~\citep{Kim20,Spiessens00} often exploited to simplify group sequential analyses since, for $j<j'$, the distribution of $S_{Bg1j'}$ differs in general from that of $S_{Bg1j}$, thus so do the distributions of $\widehat{\mu}_{A0j}(S_{Bg1j'})$ and $\widehat{\mu}_{A0j}(S_{Bg1j})$, thus
$\cov\left( \widehat{\mu}_{A0j}(S_{Bg1j}),  \widehat{\mu}_{A0j'}(S_{Bg1j'})\right)\ne \var\left( \widehat{\mu}_{A0j}(S_{Bg1j})\right)$ in general, and so
$\cov(\wDelt_{Ej},\wDelt_{Ej'})\ne \var\left( \wDelt_{Ej}\right).$
In spite of this, in Section~\ref{sec:gs.no.fut} we provide a way to compute stopping boundaries for $\jbwhat{\bm{\Delta}}_E$ utilizing the limiting  normal distribution and the covariance structure~\eqref{cov.Dj.Dj'.stat}, which can be implemented by Monte Carlo simulation.

\subsection{Group sequential tests without futility stopping}\label{sec:gs.no.fut}

Replacing the variances in \eqref{var.Dh.j} by the corresponding sample variances, for $j\in[J]$ define the variance estimator
\begin{equation*}
\jbwhat{\sigma}_{jj}=\sum_{g=0}^1 \frac{1}{n_{Bg}}\left\{ \frac{1}{n_{Bg}}\sum_{i=1}^{n_{Bg}} \left(\widehat{\mu}_{A0j}(S_{Bgij})\right)^2 - \left(\frac{1}{n_{Bg}}\sum_{i=1}^{n_{Bg}} \widehat{\mu}_{A0j}(S_{Bgij})\right)^2\right\} 
\end{equation*} of $\sigma_{jj}$.  We consider the group sequential procedure that stops and rejects the null hypothesis~\eqref{hoe_j} at the first analysis~$j\in[J]$ such that
\begin{equation}\label{gs.stop.rule}
\left|W_{Ej}\right|\ge b_j,\qmq{where} W_{Ej}:=\wDelt_{Ej}/\sqrt{\jbwhat{\sigma}_{jj}},
\end{equation} and otherwise terminates at the $J$th analysis and declares failure to reject the null hypothesis. The statistics~$W_{E1},W_{E2}, \ldots, W_{EJ}$ extend the univariate statistic, $W_E$, to the group sequential setting. In \eqref{gs.stop.rule}, the $b_j$ are pre-determined stopping boundaries; in general the $b_j$ will also depend on $J$ but we suppress this to simplify the notation. This test structure does not permit stopping before the $J$th analysis to declare failure to reject the null hypothesis, known as futility stopping, and in Section~\ref{sec:fut} we discuss a modification of this procedure that allows futility stopping.

To compute the stopping boundaries~$b_j$ to achieve approximately a prescribed type~I error probability~$\alpha$, we consider a set up that includes the popular \citet{pocock1977group}, O'Brien-Fleming~\citeyearpar{OBrien79}, and Wang-Tsiatis~\citeyearpar{Wang87} ``power family'' type boundaries, although a similar approach can be used for other families of group sequential boundaries \citep[see][]{Jennison00}. For these families of stopping boundaries,  the $b_j$ in \eqref{gs.stop.rule} are of the form 
\begin{equation}\label{cj.cdj}
b_j=b\cdot \wtilde{b}_j,
\end{equation} where $\wtilde{b}_j$ is a known function of $j$ (and $J$) and $b$ is a constant determined to satisfy the type~I error probability constraint. For example, for $j\in[J]$, the Pocock boundaries use $\wtilde{b}_j=1$, the O'Brien-Fleming boundaries use $\wtilde{b}_j=\sqrt{J/j}$, and the Wang-Tsiatis boundaries use $\wtilde{b}_j=(j/J)^{\delta-1/2}$ for a chosen value of the power parameter~$\delta$. The latter includes the Pocock ($\delta=1/2$) and O'Brien-Fleming ($\delta=0$) boundaries as special cases.  These $\wtilde{b}_j$ are intended for equally space analyses, i.e., analysis time points satisfy $t_j/t_J = j/J$. For unequally space analyses, these functions of $j/J$ should be replaced by $t_j/t_J$.


The variance estimates~$\jbwhat{\sigma}_{jj}$ used in $W_{Ej}$ in \eqref{gs.stop.rule} utilize the Study~B surrogate data~$\{S_{Bgij}:\; i=1,...,n_{Bg},\; g=0,1\}$ data from the current analysis~$j$.  However, at the design phase of the trial when the stopping boundaries~$b_j$ must be calculated, this future data will not be available (see Section \ref{discussion} for a potential adaptive approach). The normal limit~\eqref{D.hat.lim} still holds when the covariances in \eqref{cov.Dj.Dj'.stat} are replaced by the consistent sample covariance estimates, and under Assumption~(C6) the estimators
\begin{equation}\label{cgjj.cov}
c_{gjj'}:=\frac{1}{n_{Ag}} \sum_{i=1}^{n_{Ag}} \widehat{\mu}_{A0j}(S_{Agij}) \widehat{\mu}_{A0j'}(S_{Agij'})
-\left(\frac{1}{n_{Ag}} \sum_{i=1}^{n_{Ag}} \widehat{\mu}_{A0j}(S_{Agij}) \right) \left(\frac{1}{n_{Ag}} \sum_{i=1}^{n_{Ag}} \widehat{\mu}_{A0j'}(S_{Agij'}) \right),
\end{equation} $g=0,1$, are consistent estimators of the covariances~$\cov\left( \widehat{\mu}_{A0j}(S_{Bg1j}),  \widehat{\mu}_{A0j'}(S_{Bg1j'})\right)$ in \eqref{cov.Dj.Dj'.stat}. In \eqref{cgjj.cov} the Study~B surrogate data has been replaced  by that from Study~A, and therefore \eqref{cgjj.cov} can be used at the design phase to calculate the stopping boundaries of the test.  Letting 
\begin{equation*}
\wtilde{\sigma}_{jj'}:= \frac{c_{0jj'}}{n_{B0}}+ \frac{c_{1jj'}}{n_{B1}}  
\end{equation*} for $j,j'\in [J]$, let
$\wtilde{\bm{\Sigma}}$ and $\wtilde{\bm{D}}$ denote the corresponding matrices with the $\sigma_{jj'}$ replaced by $\wtilde{\sigma}_{jj'}$, i.e., $\wtilde{\bm{D}}$ is the $J\times J$ diagonal matrix with entries $\{\wtilde{\sigma}_{jj}^{-1/2}\}$ and $\wtilde{\bm{\Sigma}}$ is the $J\times J$ matrix with $(j,j')$ entry equal to $\wtilde{\sigma}_{jj'}/\sqrt{\wtilde{\sigma}_{jj}\wtilde{\sigma}_{j'j'}}$ and diagonal entries equal to $1$. Then
\vspace{-5mm}
\begin{equation}\label{D.hats.lim}
    \wtilde{\bm{\Sigma}}^{-1/2} \wtilde{\bm{D}}\jbwhat{\bm{\Delta}}_E\To N_J(\bm{0}, \bm{I})
\end{equation}
\vspace{-15mm}

\noindent under the null. The $W_{Ej}$ in \eqref{gs.stop.rule} are the entries of the vector~$\jbwhat{\bm{D}} \jbwhat{\bm{\Delta}}_E$, where $\jbwhat{\bm{D}}$ is the diagonal matrix with entries~$\{\jbwhat{\sigma}_{jj}^{-1/2}\}$. A normal limit analogous to \eqref{D.hat.lim} and \eqref{D.hats.lim} holds for $\jbwhat{\bm{D}} \jbwhat{\bm{\Delta}}_E$:
\vspace{-5mm}
\begin{equation}\label{lim.actual.stat}
    \jbwhat{\bm{\Sigma}}^{-1/2} \jbwhat{\bm{D}}\jbwhat{\bm{\Delta}}_E\To N_J(\bm{0}, \bm{I})
\end{equation} 
\vspace{-15mm}

\noindent under the null, where $\jbwhat{\bm{\Sigma}}$ is the variance-covariance matrix of $\jbwhat{\bm{D}} \jbwhat{\bm{\Delta}}_E$, of which $\wtilde{\bm{\Sigma}}$ is a consistent estimator. Since the vector~$\jbwhat{\bm{D}} \jbwhat{\bm{\Delta}}_E$ of the $W_{Ej}$ has the same asymptotic limit under the null as  $\wtilde{\bm{D}}\jbwhat{\bm{\Delta}}_E$ when both are suitably standardized, the event~\eqref{gs.stop.rule} has the same asymptotic probability under the null as the event
\vspace{-5mm}
\begin{equation}\label{asymp.rej.event}
\left|X_j\right|\ge b_j,\qmq{where} \bm{X}=(X_1,\ldots,X_J)^T:=\wtilde{\bm{\Sigma}}^{1/2}\bm{Z}\qmq{and} \bm{Z}\sim N_J(\bm{0},\bm{I}).
\end{equation}
\vspace{-15mm}

\noindent For any stopping boundaries~$b_j$ of the form~\eqref{cj.cdj},  moving $\wtilde{b}_j$ to the other side of the inequality, \eqref{asymp.rej.event} becomes $\left|X_j\right|/\wtilde{b}_j \ge b.$
From this we see that choosing $b$ to be the upper-$\alpha$ quantile of the distribution of 
\begin{equation}\label{max.Z.d}
\max_{j\in [J]} \left|X_j\right|/\wtilde{b}_j
\end{equation}  makes the test~\eqref{gs.stop.rule} approximately level-$\alpha$ . 

Since, in general, the elements~$X_j$ of $\bm{X}=\wtilde{\bm{\Sigma}}^{1/2}\bm{Z}$ in \eqref{max.Z.d} are not independent, standard tools from order statistics or extreme value theory cannot be used to calculate \eqref{max.Z.d}. However, it can be quickly and accurately computed using Monte Carlo by simulating a large number~$B$ of independent multivariate standard normals~$\bm{Z}^{(1)},\ldots, \bm{Z}^{(B)}\sim N_J(\bm{0},\bm{I})$, setting  
\vspace*{-5mm}
$$\bm{X}^{(k)} = (X_1^{(k)},X_2^{(k)},\ldots,X_J^{(k)})^T:= \wtilde{\bm{\Sigma}}^{1/2}\bm{Z}^{(k)}$$ 
\vspace*{-15mm}

\noindent for all $k\in[B]$, and taking $b$ to be the upper-$\alpha$ sample quantile of 
\vspace{-5mm}
$$\max_{j\in [J]} \left|X_j^{(1)}\right|/\wtilde{b}_j,\quad \max_{j\in [J]} \left|X_j^{(2)}\right|/\wtilde{b}_j,\quad \ldots,\quad  \max_{j\in [J]} \left|X_j^{(B)}\right|/\wtilde{b}_j.$$

\vspace*{-7mm}

\subsection{Group sequential tests with futility stopping}\label{sec:fut}

The distribution theory above makes possible a variety of modifications to the group sequential stopping rule defined above to allow futility stopping, i.e., early stopping before the $J$th stage to declare failure to reject the null.  In this section we describe one way of doing this with a flexible family of tests known as the power family of two-sided inner wedge tests; these are generalizations of stopping rules of \citet{Pampallona94} due to \citet{Jennison01}. Let $W_{Ej}$, $j\in[J]$, be as in \eqref{gs.stop.rule}. Given stopping boundaries $\{(a_j, b_j)\}_{j\in[J]}$ with 
\vspace{-5mm}
\begin{equation}\label{fut.bdr.restr}
    \mbox{$0\le a_j< b_j$ for all $j\in[J-1]$ and $a_J=b_J$,}
\end{equation}
\vspace{-15mm}

\noindent a general form of a two-sided test with an inner wedge for futility stopping stops at the first analysis $j\in[J]$ such that
\vspace{-5mm}
\begin{equation}  \label{stop.rule.inner}
\left| W_{Ej} \right|\not\in[a_j,b_j),
\end{equation} rejecting $H_E$ if the outer boundary is crossed ~$|W_{Ej}|\ge b_j$, or declaring failure to reject the null if the statistic enters the  ``inner wedge'' $|W_{Ej}|<a_j$. By virtue of the restriction~$a_J=b_J$, the test is guaranteed to stop at the $J$th stage if it has not stopped prior to that.

The power family inner wedge tests \citep[see][p.~118]{Jennison00} utilize boundaries of the form
\vspace*{-6mm}
\begin{align}
b_j&=b(j/J)^{\delta-1/2}\qmq{for} j\in[J],\label{bj.in.wdg}\\
a_j&=\begin{cases}
(a+b)(j/J)^{1/2}-a(j/J)^{\delta-1/2}\qmq{for} j_0\le j\le J,\\
0\qmq{for} 1\le j< j_0,
\end{cases}\label{aj.in.wdg}
\end{align} with constants~$a,b,\delta$, and $j_0\in[J]$, the latter denoting the earliest analysis at which it is desired to allow futility stopping. Choosing
\vspace*{-4mm}
\begin{equation}\label{abd.range}
b>0,\quad \delta\le 1,\qmq{and} a\in \left(-b,\frac{b}{(J/j_0)^{1-\delta}-1}\right)
\end{equation}
\vspace*{-12mm} 

\noindent guarantees that the resulting boundaries~\eqref{bj.in.wdg}-\eqref{aj.in.wdg} will satisfy \eqref{fut.bdr.restr}. In Web Appendix C, we provide a method for calculating the constants~$a,b$ in the boundaries~\eqref{bj.in.wdg}-\eqref{aj.in.wdg} to achieve an approximately level-$\alpha$ test.

\vspace*{-6mm}
\section{Assumptions and Empirical Assessment \label{assumptions}}
We assume the following for the surrogate marker at each time $t_j$:

\begin{enumerate}[itemsep=2pt]
\item[] (C1) $E(Y_L^{(0)}  | S_{Lj}^{(0)}=s)$ is a monotone increasing function of $s$ for $L=A,B;$
\item[] (C2) $E(Y_L^{(1)} | S_{Lj}^{(1)}=s) \ge E(Y_L^{(0)} | S_{Lj}^{(0)}=s) $ for all $s$ for $L=A,B;$
\item[] (C3) $P(S_{Lj}^{(1)} >s) \ge P(S_{Lj}^{(0)}  >s)$ for all $s$ for $L=A,B;$

\end{enumerate}
Assumptions (C1)-(C3) are not unique to our approach and parallel those required in general in surrogate marker research \citep{wang2002measure,parast2016nonparametric,wu2011sufficient,chen2007criteria,vanderweele2013surrogate}.  Assumption (C1) implies that the surrogate marker is positively related to the time of the primary outcome, (C2) implies that there is a non-negative residual treatment effect beyond that on the surrogate marker, and (C3) implies that there is a non-negative treatment effect on the surrogate marker. 

Our testing approach requires the following additional set of assumptions for the surrogate marker at each time $t_j$:
\begin{enumerate}[itemsep=2pt]
\item[] (C4) The surrogate captures a reasonably large proportion of the treatment effect on the primary outcome, described below;
\item[] (C5) The support of $S_{Bj} \supzero$ and $S_{Bj} \supone$ must be contained within the support of $S_{Aj} \supzero$;
\item[] (C6) $E(Y_A^{(0)} | S_{Aj}^{(0)}=s) = E(Y_B^{(0)} | S_{Bj}^{(0)}=s)  $ for all $s$;
\item[] (C7) $Y_L \supone \perp S_{Lj} \supzero | S_{Lj} \supone$ and $Y_L \supzero \perp S_{Lj} \supone | S_{Lj} \supzero$ for $L=A,B$.
\end{enumerate}
\noindent 
While we don't require a perfect surrogate, we do need the surrogate to be useful in some sense of surrogacy where in Assumption (C4) we measure surrogacy using the proportion of the treatment effect on the primary outcome that is explained by the treatment effect on the surrogate marker \citep{freedman1992statistical,wang2002measure}. In practice, one would want this proportion to be above some pre-specified threshold such as 0.50 \citep{lin1997estimating,elliott2023surrogate,freedman1992statistical}. If the surrogate is weak in the sense of explaining a low proportion of the treatment effect, we would expect to have lower power to test for a treatment effect on the primary outcome. In this paper, we assume that surrogate marker validation has been previously completed and there is some level of agreement clinically and statistically that the $S$ being considered is a reasonable surrogate.  Assumption (C5) is needed for kernel estimation. Assumptions (C6) means that in the control groups, the two studies share the same conditional expectations for $Y \supzero$ given $S_j \supzero$. Though this is certainly a strong assumption, some strong conditions about the transportability between Study A and Study B are necessary in order to borrow information from Study A. If no assumptions about transportability can be made, it is likely not feasible to borrow information from Study A. This assumption is at least only specific to transportability of the control group conditional means and is likely reasonable when the control groups are, for example, usual care or placebo. Assumption (C7) is needed for identifiability.

It is reasonable to ask whether these assumptions hold and what happens if they do not hold. These questions have largely not been addressed in existing work with the exception of recent work by \citet{elliott2015surrogacy} and \citet{shafie2023incorporating} in a meta-analytic framework. Assumptions (C6) and (C7) generally cannot be empirically explored with observed data because we do not have the needed quantities i.e., we only have $Y^{(g)}$ and $S^{(g)}$ for an individual if they were in group $g$, and we do not have any primary outcome measurements, $Y$, from Study B. However, Assumptions (C1)-(C5) \textit{can} be empirically explored with available data to some extent. Assumptions (C1)-(C2) can be explored for Study A, but not for Study B (because we do not have primary outcome information); (C3) can be examined in both studies. Assumption (C4) can be explored using only Study A data, while Assumption (C5) can be explored in both studies because it involves only the surrogate marker in both studies. We illustrate assessment of Assumptions (C1)-(C5) in our data application in Section \ref{example} and specifically point out the result of misuse of our method if it is used when an assumption is violated. 

\vspace*{-5mm}
 \section{Simulation Studies}\label{sec:sim}
 In this section we present simulation studies of the performance of the group sequential procedures described in Sections~\ref{sec:gs.no.fut} and \ref{sec:fut}, and compare them with a fixed sample procedure and two naive group sequential procedures. The simulations are in two settings, Setting~1 in which there is no treatment effect and Setting~2 in which there is a treatment effect. The exact data-generating mechanisms for these settings are provided in Web Appendix D. In both settings, the surrogate marker is possibly measured at $J=8$ times points~$t_j = 1,2,\ldots, 8$ and the primary outcome is measured at $t_8=8$. Data generation (for Study B) and testing procedures are replicated $1,000$ times in each setting to estimate the expected stopping time~$E(T)$, where $T$ is the time point at which the procedure terminates and rejects or declares failure to reject~$H_E$, and the probability~$P(\mbox{Rej.}\; H_E)$ of the procedure rejecting the null. The results are summarized in Tables~\ref{tab:gs.no.fut} and \ref{tab:gs.fut} which  contain the operating characteristics of the proposed group sequential procedures without and with futility stopping, respectively. Both tables include the fixed sample size test, which rejects $H_E$ if and only if \eqref{gs.stop.rule} holds for $j=J=8$. The tables also include two naive group sequential procedures, denoted by ``GS unadjusted'' and ``GS Bonferonni,'' which stop and reject $H_E$ at the earliest analysis~$j$ such that \eqref{gs.stop.rule} occurs, with $b_1=\ldots=b_J=z_{\alpha/2}$, the upper $\alpha/2$ standard normal quantile, in the unadjusted case, and with $b_1=\ldots=b_J=z_{\alpha/2}/8$ in the Bonferroni-adjusted case. Both of these procedures declare failure to reject the null at the $J=8$th analysis if \eqref{gs.stop.rule} does not occur for any $j\in[J]=[8]$.

Table~\ref{tab:gs.no.fut} shows the performance of the group sequential procedures described in Section~\ref{sec:gs.no.fut} with Pocock, O'Brien-Fleming, and Wang-Tsiatis power family boundaries (hereafter abbreviated P, OF, and WT). These are given by stopping rule~\eqref{gs.stop.rule} with boundaries of the form~\eqref{cj.cdj} with $\wtilde{b}_j=1$, $\sqrt{J/j}$, and $(j/J)^{\delta-1/2}$ with $\delta=.4$ for the P, OF, and WT boundaries, respectively. The value~$b$ in \eqref{cj.cdj} was computed by Monte Carlo as described in Section~\ref{sec:gs.no.fut} yielding $b=2.497, 2.144$, and $2.347$, respectively.  In Setting~1, there is very little savings in expected stopping time~$E(T)$ below the maximum value $T=J=8$, reflecting that the null is true and none of these procedures in this setting allow early stopping to declare failure to reject the null. The P, OF, and WT procedures maintain a type~I error probability close to the nominal level of $\alpha=.05$ while the GS unadjusted procedure has a type~I error probability that is more than three times $\alpha=.05$. In contrast, the GS Bonferroni stopping value of $z_{\alpha/2}/8 = 2.734$, which is quite conservative, results in a type~I error probability of $.021$. In Setting~2, the P, OF, and WT procedures provide roughly 1-2 analyses of savings on average over the fixed sample test, which always takes $J=8$ analyses. The P procedure is the most aggressive in early stopping, with the OF procedure the least, and the WT procedure in the middle.  The GS unadjusted procedure has the smallest expected stopping time at $5.103$, but this is due to its extreme type~I error probability exceedance in Setting~1. In Setting~2, the powers of the P, OF, and WT procedures are substantially higher than the GS Bonferroni procedure, and comparable to but slightly less than the fixed sample procedure despite their savings in expected stopping time.

Table~\ref{tab:gs.fut} shows the performance of the group sequential procedures that incorporate futility stopping described in Section~\ref{sec:fut}, given by the stopping rule~\eqref{stop.rule.inner} with boundaries of the form~\eqref{bj.in.wdg}-\eqref{aj.in.wdg} with $\delta=1/2$, $0$, and $.4$, respectively. We refer to these in the table as Pocock, O'Brien-Fleming, and Wang-Tsiatis boundaries (and again abbreviate them here as P, OF, and WT), respectively, because they utilize the same $\delta$ values in \eqref{bj.in.wdg}-\eqref{aj.in.wdg}, although these inner wedge stopping rules were not part of those authors' original proposals.
 For these procedures, futility stopping begins at the $j_0=4$th analysis. The values~$(a,b)$ in \eqref{bj.in.wdg}-\eqref{aj.in.wdg} were computed by Monte Carlo as described in Section~\ref{sec:fut} with $\alpha_0=(j_0/J)\alpha = \alpha/2=.025$ yielding $(a,b)=(2.455, 4.4)$, $(2.008, 1.7)$, and $(2.345,3.3)$, respectively. The fixed sample and unadjusted and Bonferroni GS procedures are the same as in Table~\ref{tab:gs.no.fut}. The P, OF, and WT procedures with futility stopping show more savings in $E(T)$ in both Settings~1 and 2, with the P procedure being the most aggressive in early stopping, followed by the WT and OF procedures.  These show substantial savings over the fixed sample and naive GS procedures, which do not allow futility stopping, in Setting~1, and even savings over these procedures in Setting~2, with the exception of the GS unadjusted procedure whose type~I error probability is more than $3$ times $\alpha=.05$.  The power of the P, OF, and WT procedures is higher than the GS Bonferroni procedure, which is conservative in this sense of being underpowered in Setting~2 as well as having a type~I error probability of $.021$ in Setting~1.

\vspace*{-5mm}

\section{Application to AIDS Clinical Trials}\label{example}
We illustrate our proposed testing procedure using data from two randomized clinical trials from the AIDS Clinical Trial Group (ACTG) Network: ACTG 320 and ACTG 193A, where ACTG 320 is Study A and ACTG 193A is Study B. The ACTG 320 study was a randomized, double-blind trial among 1,156 HIV-infected patients that compared a two-drug regimen (two nucleoside reverse transcriptase inhibitors [NRTI]; $n=579$) with a three-drug regimen (two NRTIs plus indinavir; $n=577$) \citep{hammer1997controlled}.  The ACTG 193A study was a randomized, double-blind trial among 1,313 HIV-infected patients comparing four daily treatment regimes \citep{henry1998randomized}. We focus on comparing a two-drug regimen (zidovudine plus zalcitabine, 2 NRTIs; $n=326$, group 0) with a three-drug regimen (two NRTIs plus nevirapine; $n=330$, group 1) in this study. 

For our analysis, the primary outcome is change in RNA from baseline to 40 weeks post-baseline which was measured in Study A but was \emph{not} measured for all participants in Study B. The surrogate marker of interest is change in CD4 count from baseline to $t_j$, $0<t_j\leq 40$. That is, we aim to test for a treatment effect on change in RNA at 40 weeks in Study B (ACTG 320 Study) using the surrogate marker of change in CD4 count via our proposed group sequential testing approach. Importantly, the exact times of the CD4 measurements do not perfectly align between studies. In Study A, CD4 was measured at baseline, 4, 8, 24, and 40 weeks; in Study B, CD4 was measured at baseline, 8, 16, 24, and 40 weeks. Before we discuss how to handle this, we first assessed Assumptions (C1)-(C3) (Section \ref{assumptions}). Detailed results, provided in Web Appendix E, show that we should not borrow information about the surrogate marker at weeks 4 or 8 from Study A. Furthermore, there is no measurement available at 4 weeks in Study B. In Study B, we only consider testing at 16, 24, and 40 weeks. But, because the surrogate was not measured at 16 weeks in Study A, we construct the test statistic at 16 weeks by borrowing the conditional mean function $\widehat{\mu}_{A0}(s)$ at 24 weeks.

\begin{figure}[htbp]
\begin{center}
\includegraphics[scale=0.9]{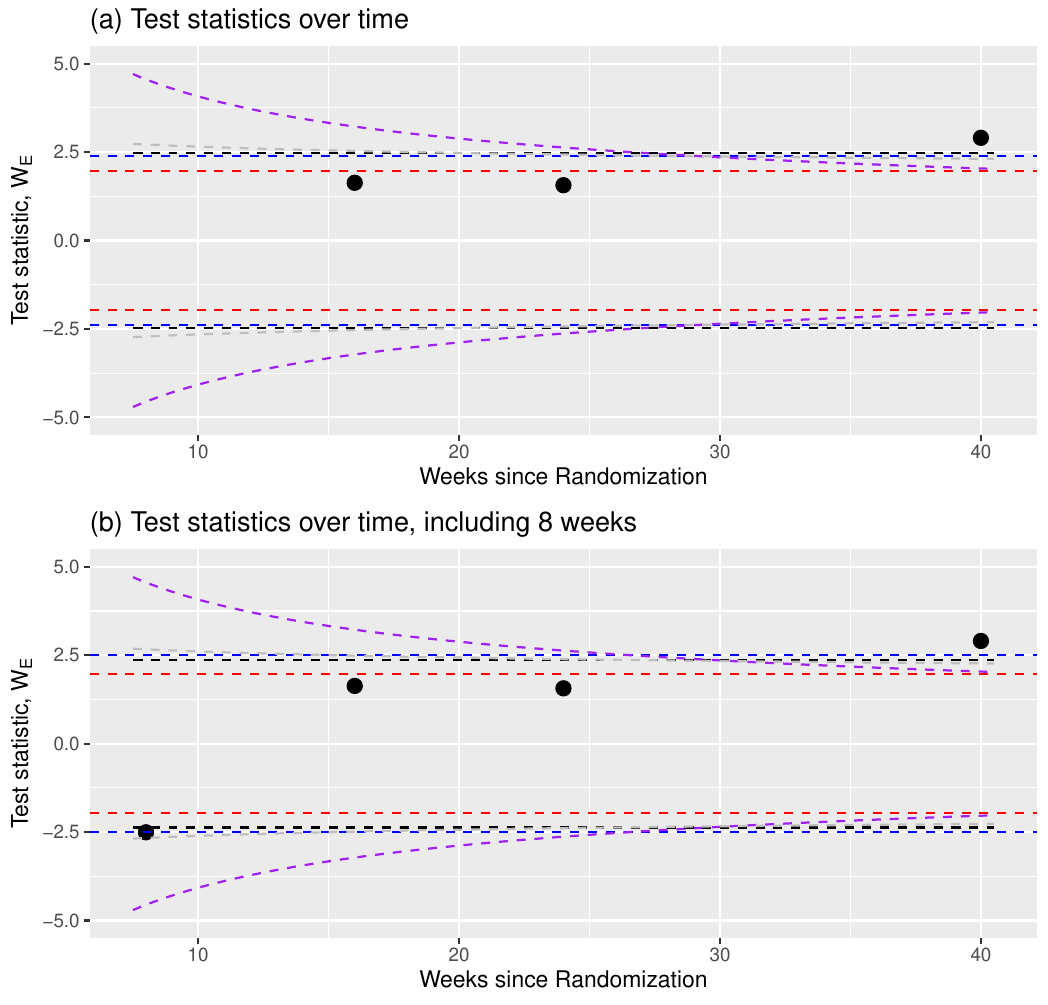}
\end{center}
\caption{AIDS clinical trial testing results where black dots indicate the estimated test statistic, $W_E$, at each time point $t_j$ for weeks 16, 24, and 40 (a), and weeks 8, 16, 24, and 40 (b); red dashed line is drawn at $\pm \Phi^{-1}(1-0.05/2)=1.96$, blue dashed line is drawn at Bonferroni boundary, purple dashed line is O'Brien-Fleming boundary, black dashed line is Pocock boundary, and grey dashed line is Wang-Tsiatis boundary.} \label{aids_plot}
\end{figure}

Figure \ref{aids_plot}(a) shows the resulting test statistics with dashed lines drawn at the naive, Bonferroni, Pocock, O'Brien-Fleming, and Wang-Tsiatis (with $\delta=.4$) boundaries as described in Section~\ref{sec:gs.no.fut} with $J=3$ analyses. The constant~$b$ in \eqref{cj.cdj} for each of these procedures was calculated by Monte Carlo as $2.370$, $2.038$, and $2.268$, respectively, via the Monte Carlo method described in Section~\ref{sec:gs.no.fut} with the quantity~$j/J$ used in the $\wtilde{b}_j$ expressions replaced by $t_j/t_J$ with $t_j=16$, $24$, and $40$ because of the unequally spaced analyses. For all boundaries, we would conclude a significant treatment effect at 40 weeks using the surrogate marker, CD4 count. To illustrate the importance of empirically examining the needed assumptions (which we do in Web Appendix E) in practice, we show in Figure \ref{aids_plot}(b), the results we \textit{would have} obtained had we included testing at 8 weeks in Study B, using the surrogate marker information at 8 weeks from Study A, even though we have empirical evidence of possible assumption violation.  Here, we would have incorrectly concluded a significant negative treatment effect using the naive, Bonferroni, and Pocock boundary, but not with the O'Brien-Fleming or and Wang-Tsiatis boundaries. In practice, it is important that the needed assumptions be examined empirically as described in Section \ref{assumptions}, to the extent possible with available data, to reduce the risk of an inappropriate conclusion (such as this one) based on the surrogate marker. 

\vspace*{-5mm}
\section{Discussion \label{discussion}}

There is a tremendous amount of hope in the promise of surrogate markers that can identify effective (and ineffective) treatments sooner.  In this paper, we offer group sequential procedures that allow for early stopping to declare efficacy of the treatment effect or early futility stopping. If the surrogate measurements over time were independent, this would be a quite straightforward problem. However, it would be unreasonable to assume such independence, at least in any clinical setting, and thus, our statistical development of appropriate procedures was complicated by and accounted for the correlation between the surrogate marker measurements over time. An R package that implements our procedures, named \texttt{SurrogateSeq}, is available at \url{https://github.com/laylaparast/SurrogateSeq}.

The simulation studies in Section~\ref{sec:sim} show that the Pocock, O'Brien-Fleming, and Wang-Tsiatis versions of the proposed group sequential procedures can be made to control the type~I error close to a nominal level~$\alpha$ using the Monte Carlo methods described in Sections~\ref{sec:gs.no.fut} and \ref{sec:fut}.  The versions of these procedures without futility stopping provide modest savings in sample size while maintaining comparable power, and the versions incorporating futility stopping provide more savings in sample size while having higher power than the Bonferroni-adjusted, naive group sequential procedure.  Our results show that the naive, unadjusted group sequential procedure is not a viable option because its type~I error probability grossly exceeds the nominal level, a well-understood phenomenon in many sequential and group sequential settings. In practice, when deciding between the Pocock, O'Brien-Fleming, and Wang-Tsiatis versions of the proposed procedures, we recommend the Pocock boundaries if aggressive early stopping or simplicity of stopping boundaries (i.e., constant) is desired, the O'Brien-Fleming boundaries if maintaining high power is a primary objective; otherwise, some instance of the Wang-Tsiatis boundaries can be chosen as a ``middle ground.''

While a major advantage of our proposed approach is the lack of reliance on parametric assumptions, we still require a number of assumptions described in Section \ref{assumptions}. These assumptions are not unique to our approach; some version of these assumptions is generally required in surrogate marker research (see \citet{vanderweele2013surrogate}). Importantly, we do not claim that these are \textit{replacing} parametric assumptions and are less strict. They are strict assumptions, and while we can empirically examine some of these with available data, we cannot examine or test Assumptions (C6) and (C7). Assumption (C6) has some parallels to assumptions used in domain adaptation \citep{kouw2019review} and Assumption (C7) has some parallels to (though is not equivalent to) the widely used assumption of no unmeasured confounding in causal inference \citep{vanderweele2011bias}. Future work to develop methods to investigate these assumptions via simulation or the construction of bounds would be extremely valuable. 

The proposed method for calculating the stopping boundaries of these procedures in Section~\ref{sec:gs.no.fut} relies on Assumption~(C6) which permits estimating the covariance structure of the Study~B test statistics using that from Study~A.  If Study~A does not provide an accurate estimate of the Study~B covariance, an alternative approach may be to compute the boundaries adaptively using a type~I error-spending function~$\alpha(t)$, increasing from $\alpha(0)=0$ to $\alpha(1)=\alpha$.  In the context of the group sequential tests in Section~\ref{sec:gs.no.fut} with stopping rule~\eqref{gs.stop.rule}, after the $(j-1)$st analysis has been completed and data from the $j$th analysis is available, the stopping boundary~$b_j$ would be computed to satisfy
\vspace*{-5mm}
\begin{equation}\label{adapt.rej.j}
P(|W_{E1}|<b_1,\ldots, |W_{Ej-1}|<b_{j-1}, |W_{Ej}|\ge b_j)=\alpha(t_j/t_J) - \alpha(t_{j-1}/t_J),
\end{equation}
 where this probability is under the null hypothesis.  The event in  \eqref{adapt.rej.j} can be simulated via Monte Carlo using a similar approach to  \eqref{asymp.rej.event} but where $\bm{Z}$ is length $j$ and $\wtilde{\bm{\Sigma}}$ is replaced by  the  $j\times j$ leading principal submatrix of $\jbwhat{\bm{\Sigma}}$, a function of the available data from the first $j$ analyses, producing length~$j$ vectors~$\bm{X}$ whose entries have the same distribution as $W_{E1}, W_{E2}, \ldots, W_{Ej}$ in \eqref{adapt.rej.j} under the null. Any of the popular methods \citep{Slud82,Lan83,Lan89} could be used for choosing the error spending function~$\alpha(t)$. 

While this work is a unique and novel contribution to the field, the practical utility of this approach depends immensely on the quality, size, and representativeness of the Study A data. Notably, our testing framework makes explicit the reliance on Study A data via the definition of $\Delta_E$ which is defined with the $\widehat{\mu}_{A0j}(s)$ term. While it is uncommon to see an estimator within a defined null hypothesis, this reflects the true use of surrogates in practice. It would be disingenuous to develop a testing procedure and asymptotic results under the assumption that $n_A \rightarrow \infty$ because it \textit{does not}. When a study team is considering using a surrogate marker to test for a treatment effect, they are using information about surrogacy from some previously completed study. The design is such that this prior study, Study A, is fixed, not random. (However, for completeness, we discuss our approach with Study A as random in Web Appendix F.) If Study A is very small or not representative or in some way not providing a reasonable estimate of  $\mu_{A0j}(s)$, there is no magic (that we know of) that would produce a legitimate testing procedure that uses the surrogate as a replacement of the primary outcome. Ultimately, this is a very difficult problem. Surrogate markers are not perfect and it is important for us to be explicit about the reliance of this testing procedure (and more generally, any testing procedure that uses a surrogate marker to replace the primary outcome) on prior data that is being used to assess surrogacy.

\vspace*{-7mm}
\section*{Acknowledgements}
This work was supported by NIDDK grant R01DK118354. We are grateful to the AIDS Clinical Trial Group (ACTG) Network for supplying the ACTG data.

\vspace*{-7mm}
\section*{Data Availability Statement}
The AIDS clinical trial data used in this paper are publicly available in our R package \texttt{SurrogateSeq}, available at \url{https://github.com/laylaparast/SurrogateSeq}. 

\vspace*{-7mm}
\section*{Supplementary Material}
Referenced Web Appendices are available with this paper at the Biometrics website. 

\vspace*{-7mm}

\clearpage

 \clearpage

\begin{table}[]
\caption{Expected stopping time $E(T)$ and probability~$P(\mbox{Rej.\ $H_E$})$ of rejecting the null hypothesis~$H_E$ for group sequential procedures without futility stopping, at most $J=8$ analyses, and nominal level~$\alpha=.05$ based on $1,000$ replications of each procedure in each setting. Standard errors~SE are in parentheses.}
\label{tab:gs.no.fut}
\vspace*{5mm}
\begin{tabular}{l||c|c||c|c}
\hline
& \multicolumn{2}{c||}{Setting 1: no treatment effect} & \multicolumn{2}{c}{Setting 2: treatment effect}\\
Test & $E(T)$ (SE) & $P(\mbox{Rej.\ $H_E$})$ (SE)                 & $E(T)$ (SE) & $P(\mbox{Rej.\ $H_E$})$ (SE) \\\hline
Fixed sample test & 8.000 (.000) & .053 (.007) & 8.000 (.000) & .822 (.012)\\
GS unadjusted & 7.314 (.060) & .151 (.011)  & 5.103 (.073) & .874 (.010)  \\
GS Bonferroni & 7.921 (.020)   & .021 (.005) & 6.694 (.056) & .586 (.016)\\
Pocock & 7.792 (.034) & .045 (.007) & 6.173 (.065) & .723 (.014) \\
O'Brien-Fleming & 7.938 (.012)  & .054 (.007) & 6.609 (.043) & .808 (.012) \\
Wang-Tsiatis ($\delta= .4$) & 7.866 (.024) & .046 (.007) & 6.210 (.058)  & .762 (.013) \\ \hline
\end{tabular}

\end{table}

\begin{table}[]
\caption{Expected stopping time $E(T)$ and probability~$P(\mbox{Rej.\ $H_E$})$ of rejecting the null hypothesis~$H_E$ for group sequential procedures with futility stopping beginning at the $j_0=4$th analysis, at most $J=8$ analyses, and nominal level~$\alpha=.05$ based on $1,000$ replications of each procedure in each setting. Standard errors~SE are in parentheses.}
\label{tab:gs.fut}
\vspace*{5mm}
\begin{tabular}{l||c|c||c|c}
\hline
& \multicolumn{2}{c||}{Setting 1: no treatment effect} & \multicolumn{2}{c}{Setting 2: treatment effect}\\
Test & $E(T)$ (SE) & $P(\mbox{Rej.\ $H_E$})$ (SE) & $E(T)$ (SE) & $P(\mbox{Rej.\ $H_E$})$ (SE) \\\hline
Fixed sample test & 8.000 (.000) & .053 (.007) & 8.000 (.000) & .822 (.012)\\
GS unadjusted & 7.314 (.060) & .151 (.011)  & 5.103 (.073) & .874 (.010)  \\
GS Bonferroni & 7.921 (.020)   & .021 (.005) & 6.694 (.056) & .586 (.016)\\
Pocock & 4.756 (.033) & .051 (.007) & 5.572 (.056) & .662 (.015) \\
O'Brien-Fleming & 5.340 (.032)  & .053 (.007) & 5.619 (.040) & .706 (.014) \\
Wang-Tsiatis ($\delta= .4$) & 5.052 (.033) & .052 (.007) & 5.576 (.051)  & .682 (.015)\\ \hline
\end{tabular}

\end{table}

 \clearpage

\end{document}